\documentclass[dvips]{article}

\usepackage{icrc2011}
\usepackage{amsmath}
\usepackage{textcomp}

\title{Seasonal modulation in the Borexino cosmic muon signal}

\newcommand{\etal}{\MakeLowercase{\textit{et al. }}} 
\shorttitle{D.D'Angelo \etal Seasonal modulation in the Borexino cosmic muon signal}

\authors{Davide D'Angelo$^{1}$ for the Borexino collaboration}

\afiliations{$^1$Universit\`a degli Studi di Milano and I.N.F.N. sez. Milano, via Celoria 16 - Milano (Italy)}

\email{davide.dangelo@mi.infn.it}

\abstract{
Borexino is an organic liquid scintillator detector located in the underground Gran Sasso National Laboratory (Italy).
It is  devoted mainly to the real time spectroscopy of low energy solar neutrinos via the elastic scattering on electrons in the target mass. 
The data taking campaign started in 2007 and led to key measurements of $^{7}$Be and $^{8}$B solar neutrinos as well as antineutrinos from the earth (geo-neutrinos) and from nuclear power reactors.
Borexino is also a powerful tool for the study of cosmic muons that penetrate the Gran Sasso rock coverage and thereby induced signals such as neutrons and radioactive isotopes which are today of critical importance for upcoming dark matter and neutrino physics experiments. 
Having reached 4y of continuous data taking we analyze here the muon signal and its possible modulation.
The muon flux is measured to be $(3.41\pm0.01)\cdot10^{-4}$m$^{-2}$s$^{-1}$. 
A modulation of this signal with a yearly period is observed with an amplitude of (1.29$\pm$0.07)\% and a phase of (179$\pm$6) d, corresponding to June $28^{th}$.
Muon rate fluctuations are compared to fluctuations in the atmospheric temperature on a daily base, exploiting the most complete atmospheric data and models available.
The distributions are shown to be positively correlated and the effective temperature coefficient is measured to be $\alpha_T = 0.93 \pm 0.04$. 
This result is in good agreement with the expectations of the kaon-inclusive model at the laboratory site
and represents an improvement over previous measurements performed at the same depth.
}


\begin{document}
\maketitle

{\bf I. Introduction.} The flux of cosmic muons detected deep underground shows variations which are in first approximation seasonal. 
The effect is known and studied since many decades \cite{bar52}.
At Gran Sasso National Laboratory in central Italy the rock coverage is about 3800 m w.e. and the expected amplitude of the modulation is $\sim$1.5\%. 
Borexino is a scintillator detector with an active mass for muon detection of 1.33 kt and, being spherical, its acceptance is independent of the angle of the incoming muons.
It therefore plays a key role in measuring the cosmic muon flux and its modulation with reduced systematics.
Moreover as air temperature data is available from weather forecast web services, the correlation with the muon flux can be investigated and the effective temperature coefficient can be determined.
Such temperature coefficient, with a larger exposure, can in future be used to determine indirectly the K/$\pi$ ratio in the interaction of primary cosmic rays in the atmosphere, probing a complementary energy region compared with existing accelerator experiments.

\vspace{2mm}

{\bf II. Borexino Detector.} The Borexino detector was designed to have very low intrinsic background.  
The central scintillation volume, 278\,t
of ultra-pure PC (pseudocumene) doped with 1.5\,g/l of the fluor PPO (2,5-diphenyloxazole), is contained in a spherical
Inner Vessel (IV), 8.5\,m in diameter, made of 125\,\textmu\ m thick nylon. 
It is shielded by two buffer layers consisting of PC and a small amount of the light quencher DMP (dimethylphthalate). 
The surrounding Stainless Steel Sphere (SSS) of 13.7\,m diameter holds 2212 inward-facing 8'' photomultiplier tubes (PMT) 
that detect scintillation light from the central region. 
All these components form the Inner Detector (ID) \cite{bx08det}.

Though Borexino is located deeply underground, in Hall C of the Gran Sasso Laboratory (LNGS),
the residual muon flux is  $\sim$\,1.2\,\textmu/m$^2$/h,
still too large for neutrino measurement, so the muons must be individually tagged. 
To accomplish this task the ID is surrounded by a powerful muon detector \cite{bx11mv}.
It is composed by a high domed steel tank of 18\,m diameter and 16.9\,m height filled
with 2\,100\,t of ultra-pure water and instrumented with 208 PMTs which detect the muon \v{C}erenkov emission. 
The Water Tank also serves as additional passive shielding against external radiation. 
This system is called the Outer Detector (OD).

\vspace{2mm}

{\bf III. Cosmic Muon Flux.} This analysis is based on the first 4 years of Borexino data, taken between May 16$^{th}$ 2007 and May 15$^{th}$ 2011, 
with the exclusion of calibration data, data not passing the validation procedure and data for which the OD was not functioning properly.
Events in coincidence with the spills of the Cern-to-GranSasso neutrino beam are discarded (details in \cite{bx11mv}). 
The remaining data set shows no prolonged or unevenly distributed off time.
This analysis is based only on muon events that triggered both ID and OD.
The total resulting exposure is $\sim$ 1.41 $\cdot$ $10^{6}$ t$\cdot$d and includes a sample of $\sim 4.6 \cdot 10^{6}$ muons.

We have measured the muon rate through the ID using different strategies at our disposal and achieved identical results. 
The overall detector's efficiency is 99.992\%.
In \cite{bx11mv} are reported details on the muon tagging methods and on how the efficiencies have been evaluated.
The average muon rate is (4310$\pm$10) counts per day, where the statistical error is negligible and the systematic error reflects the uncertainty in the efficiency and possible threshold effects.
The rate corresponds to a cosmic muon flux of $(3.41\pm0.01)\cdot10^{-4}$m$^{-2}$s$^{-1}$, taking into account also the uncertainty in the SSS radius.

This is the first measurement available for Hall C so far and the first obtained with a spherical detector at LNGS:
existing measurements were obtained with detectors whose acceptance strongly depended on the muon incidence angle and are therefore affected by larger systematics. 
They have been performed by LVD in Hall A (\cite{sel09}) and by MACRO in Hall B (\cite{macro95}) and are respectively $(3.31\pm0.03)\cdot 10^{-4}$m$^{-2}$s$^{-1}$ and $(3.22\pm0.08)\cdot10^{-4}$m$^{-2}$s$^{-1}$.

\vspace{2mm}

{\bf IV. Flux Modulation.} Muons observed in underground sites arise mostly from the decay of pions and kaons produced by primary cosmic ray particles interacting with nuclei in the atmosphere\cite{gai90}. 
Only mesons decaying before further interaction produce muons energetic enough to traverse the rock coverage of an underground site.
Air temperature increases during summer, leading to an expansion of the traversed medium, which in turn increases the fraction of such mesons.
The following formula is generally used to relate the muon intensity variations to the atmospheric temperature fluctuations:
\begin{equation}
\frac{\Delta I_{\mu}}{I^0_{\mu}} = \int_{0}^{\infty}dX\alpha(X) \frac{\Delta T(X)}{T^0 (X)}
\label{eq:seasonal}
\end{equation}
where $I^0_{\mu} = I_{\mu}(T_{0},E > E_{thr})$ is the differential
muon intensity integrated from the energy threshold
($E_{thr}$ $\sim$ 1.8 TeV)\cite{gra10} to infinity, assuming the atmosphere
is isothermal at temperature $T^{0}$, and $\Delta I_{\mu}$ are fluctuations
about $I^0_{\mu}$; $\alpha(X)$ is the temperature coefficient that
relates fluctuations in the atmospheric temperature at
depth X, $\Delta T(X)/T^0(X)$, to the fluctuations in the
integral muon intensity; the integral extends over atmospheric
depth from the altitude of muon production to the ground. 

Other underground experiments have studied these effects, at the Gran Sasso site (MACRO \cite{macro97}, LVD \cite{sel09}) and at different underground locations( \cite{amanda}, \cite{minos10} and refs. therein). 

\begin{figure*}[!tbh]
\centering
\includegraphics[width=\textwidth]{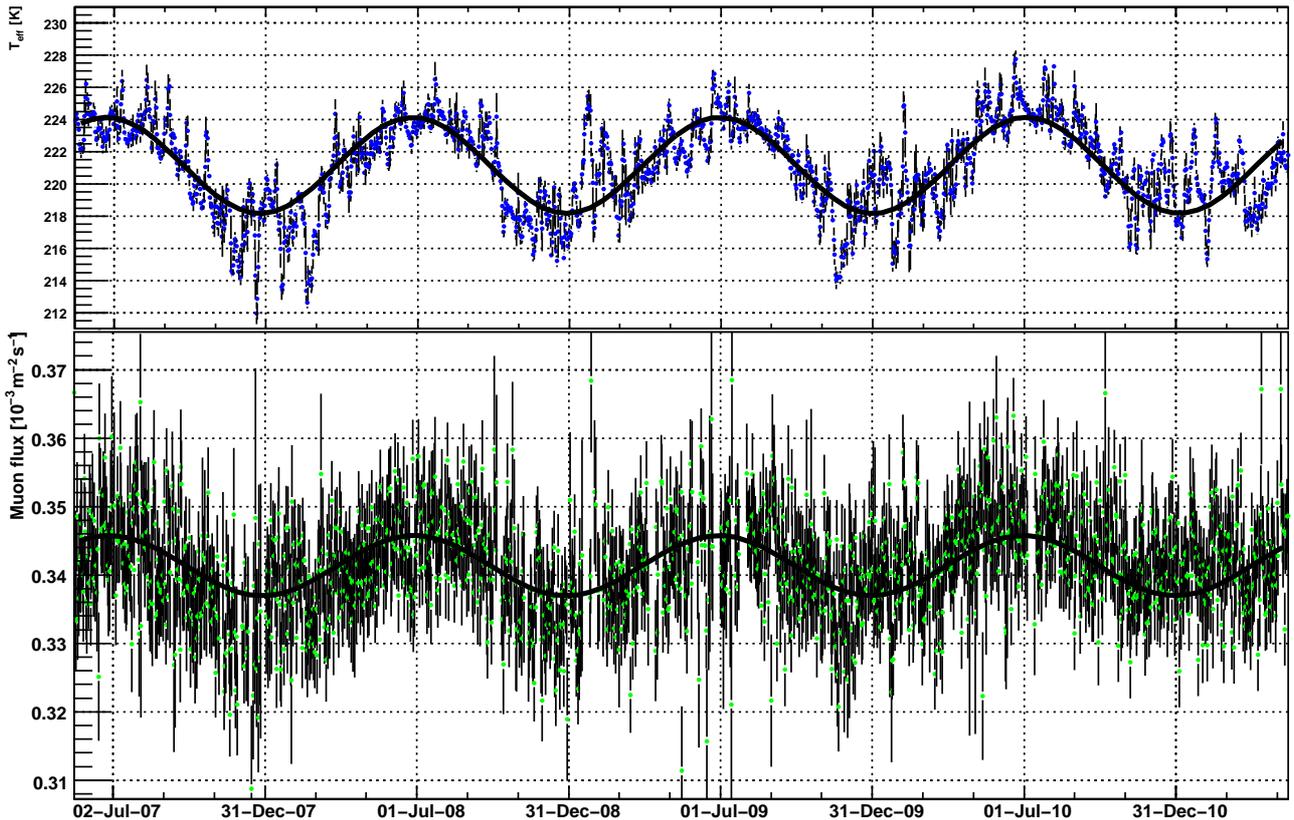}
\caption{Upper panel: effective temperature $T_{\textrm{eff}}$ computed day by day using eq. \ref{eq:teff} and averaging over the 4 daily measurements available. Lower panel: Cosmic Muon Signal. The seasonal modulation is evident. Daily binning.}
\label{fig:temp_flux}
\end{figure*}

The muon intensity measured day by day is shown in fig. \ref{fig:temp_flux} (lower panel) for the 1329 days for which valid data was available.
A modulation is clearly visible; fitting the distribution with the following function:
\begin{equation}
I_{\mu} = I_{\mu}^{0} + \delta I_{\mu} cos \left(\frac{2\pi}{T}(t-t_{0})\right)
\label{eq:seasonal_fit}
\end{equation}
we obtain an average intensity $I_{\mu}^{0}=(3.414\pm0.002)\cdot 10^{-4}$m$^{-2}$s$^{-1}$, 
a period $T=(366\pm3)$d, 
a modulation amplitude $\delta I_{\mu}=(4.4\pm0.2) \cdot 10^{-6}$m$^{-2}$s$^{-1}$ corresponding to $(1.29\pm0.07)\%$ 
and a phase $t_{0}=(179\pm6)$d; 
the $\chi^{2}$/NDF is 1558/1325.
A Lomb-Scargle analysis of the data identifies the same period.
It should be noted that due to the limited size of the detector, a day with 100\% duty cycle features a statistical error of $\sim$1.5\% (1$\sigma$), comparable to the expected modulation. 
Therefore in spite of the fair value of reduced $\chi^{2}$, we regard this fitting exercise only as a first order approximation; 
in the hypothesis, explored here, that the modulation is related to the air temperature fluctuations, the main maxima and minima can occur at different dates in successive years and short term effects are well expected to perturb the overall seasonal behavior. 

\vspace{2mm}

{\bf V. Atmospheric Model.} The atmosphere consists of many layers that vary continuously in temperature and pressure.
A possible parametrization (\cite{minos10} and with more details\cite{gra10}) considers the atmosphere as an isothermal body with an effective temperature, $T_{\textrm{eff}}$, obtained from a weighted average over atmospheric depth:
\begin{equation}
T_{\textrm{eff}} = \frac{\int_0^{\infty}dXT(X)W(X)}{\int_0^{\infty}dXW(X)}
\label{eq:teff_int}
\end{equation}
where the weight $W(X)$ reflects the temperature dependence of the production of mesons in the atmosphere and their decay into muons that can be observed at depth.

Eq. \ref{eq:seasonal} can now be written in term of the ``effective temperature coefficient'' $\alpha_T$:
\begin{equation}
\frac{\Delta I_{\mu}}{I^0_{\mu}} = \alpha_T \frac{\Delta T_{\textrm{eff}}}{T_{\textrm{eff}}}
\label{eq:var_at}
\end{equation}

The weight $W(X)$ can be written as the sum $W_{\pi}+W_{K}$, representing the contribution of pions and kaons to the overall variation in muon intensity. 
\begin{equation}
W^{\pi,K}(X) \simeq \frac{(1-X/\Lambda'_{\pi,K})^2 e^{-X/\Lambda_{\pi,K}} A^1_{\pi,K}}{\gamma+(\gamma+1)B^1_{\pi,K}K(X)(\langle E_{th}\cos\theta\rangle /\epsilon_{\pi,K})^2}
\label{eq:weights}
\end{equation}
where:
\begin{equation}
K(X) \equiv \frac{(1-X/\Lambda'_{\pi,K})^2}{(1-e^{-X/\Lambda'_{\pi,K}})\Lambda'_{\pi,K}/X}
\label{eq:KX}
\end{equation}

The parameters $A^1_{\pi,K}$ include the amount of inclusive meson production in the forward fragmentation region, masses of mesons and muons, and muon spectral index; 
the input values are $A^1_{\pi} = 1$ and $A^1_K = 0.38 \cdot r_{K/\pi}$, where $r_{K/\pi}$ is the $K/\pi$ ratio. 
The parameters $B^1_{\pi,K}$ reflect the relative atmospheric attenuation of mesons; 
the threshold energy, $E_{th}$, is the energy required for a muon to survive to a particular depth; 
the attenuation lengths for the cosmic ray primaries, pions and kaons are $\Lambda_N$, $\Lambda_{\pi}$ and $\Lambda_K$ respectively with $1/\Lambda'_{\pi,K}= 1/\Lambda_N-1/\Lambda_{\pi,K}$. 
The muon spectral index is given by $\gamma$. 
The meson critical energy, $\epsilon_{\pi,K}$, is the meson energy for which decay and interaction have an equal probability. 
The value of $\langle E_{th} \cos \theta \rangle$ used here is the median of the distribution. 
The values for these parameters can be found in tab. 1 of \cite{minos10}, with the exception of $\langle E_{th} \cos \theta \rangle$ which is site dependent and is found by MC simulations.
At LNGS $\langle E_{th} \cos \theta \rangle$ = 1.833 TeV according to \cite{gra10}.
The dependency of $W(X)$ on $T_{\textrm{eff}}$ is however moderate.

Since the temperature is measured at discrete atmospheric levels $X_n$, eq. \ref{eq:teff_int} becomes:
\begin{equation}
T_{\textrm{eff}}\simeq\frac{\sum^N_{n=0} \Delta X_nT(X_n) (W^{\pi}_n+W^K_n)}{\sum^N_{n=0} \Delta X_n(W_n^{\pi}+W^K_n)}
\label{eq:teff}
\end{equation}
where $W^{\pi,K}_n \equiv W^{\pi,K} (X_n)$.

\begin{figure}[!thb]
\centering
\includegraphics[width=\columnwidth]{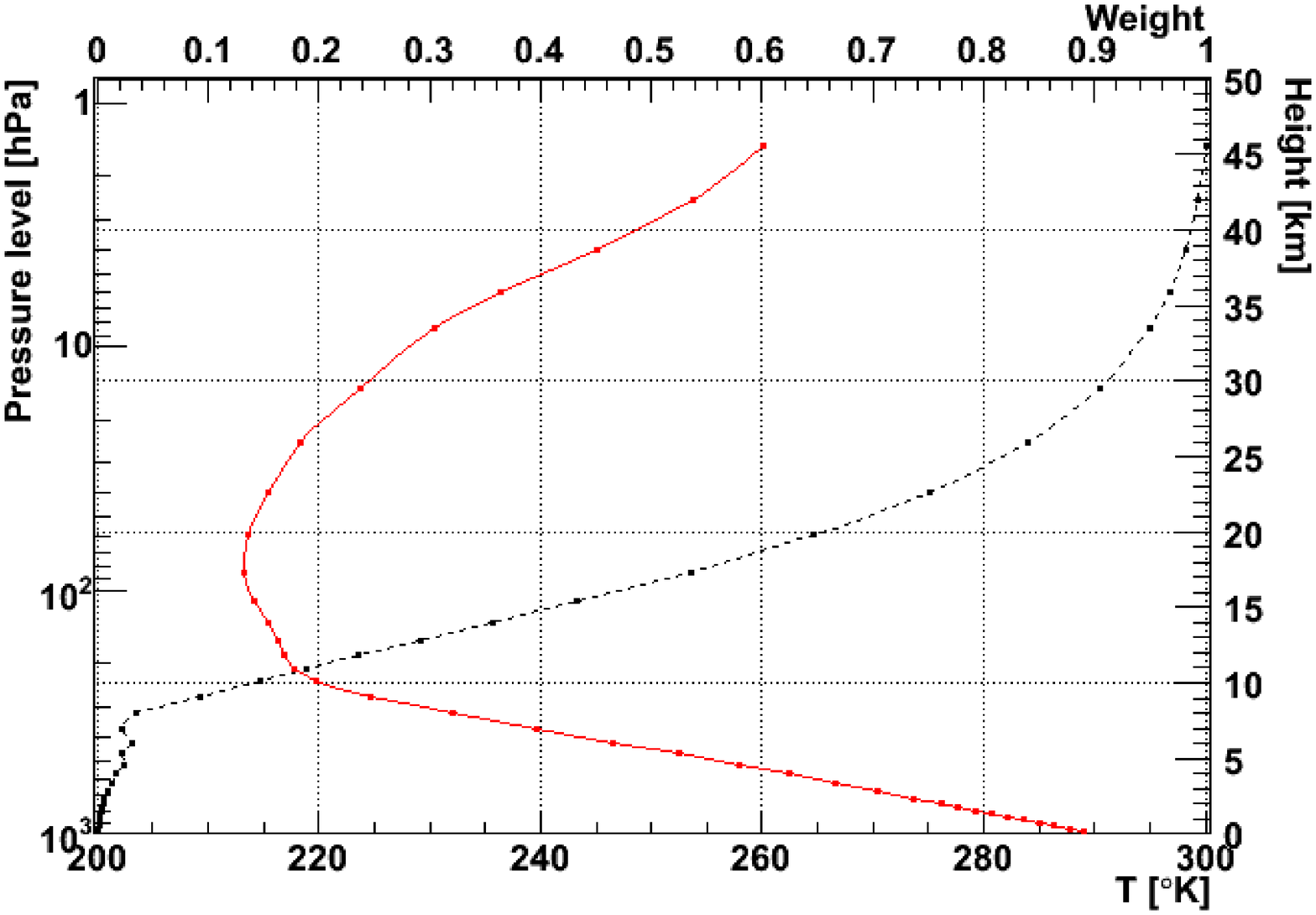}
\caption{Average temperature (solid red line) and normalized weight $W(X)$ (black dashed line) as a function of pressure levels computed at the LNGS site.} 
\label{fig:tavg_weight}
\end{figure}
 
Fig.~\ref{fig:tavg_weight} shows the temperature in the atmosphere for the LNGS site and the weights used in eq.~\ref{eq:teff} as functions of the pressure levels.
As it can be seen the higher layers of atmosphere are given a higher weight. 
Here are produced most of the muons which are energetic enough to cross the rock coverage of an underground site.
Muons produced in lower level will be in average less energetic and a larger fraction of them lies below threshold. 

\vspace{2mm}

{\bf VI. Temperature Modulation.} The temperature data was obtained from the European Center for Medium-range Weather Forecasts (ECMWF)\cite{ecmwf}
which exploits different types of observations (e.g. surface, satellite and upper air sounding) at many locations around the planet, 
and uses a global atmospheric model to interpolate to a particular location. 
In our case, the exact LNGS coordinates have been used: 13.578E, 42.454N.
Atmospheric temperature is provided by the model at 37 discrete pressure levels in the [1-1000]hPa range (1 hPa = 1.019 g/cm$^2$ ), 
four times a day at 00.00 h, 06.00 h, 12.00 h and 18.00 h. 
Based on this data set, the effective temperature $T_{\textrm{eff}}$ was calculated using eq.~\ref{eq:teff} four times a day\footnote{The analysis in \cite{macro97} and \cite{sel09} instead used data from the air soundings performed by the Areonautica Militare Italiana (AM)\cite{am} $\sim$130km from the lab. 
This data set is significantly incomplete if compared to ECMWF, although it provided us a useful cross-check.}

For the 4y period $\langle T_{\textrm{eff}} \rangle$ = 220.99K.
Fig.~\ref{fig:temp_flux} (upper panel) shows  $\langle T_{\textrm{eff}}  \rangle$ for each day obtained averaging the 4 available measurement and estimating the error from their variance.
The fit with a function analogous to eq. \ref{eq:seasonal_fit} returns $T_{\textrm{eff}}^{fit}=(221.153\pm0.007)$K, 
amplitude $(2.98\pm0.01)$K corresponding to 1.35\%, period $T=(369.2\pm0.2)\textrm{d}$ and phase $(174.0\pm0.4)\textrm{d}$. 
However here the error bars are much smaller and the $\chi^2$/NDF is very poor confirming that the sinusoidal behavior is only a first order approximation.
Aside from small scale fluctuations, additional winter maxima can be observed which can be ascribed to the known meteorological phenomenon of the Sudden Stratospheric Warmings (SSW \cite{osp09}) and whose effect is sometimes comparable in amplitude with the underlying seasonal modulation.

\begin{figure}[!thb]
\centering
\includegraphics[width=\columnwidth]{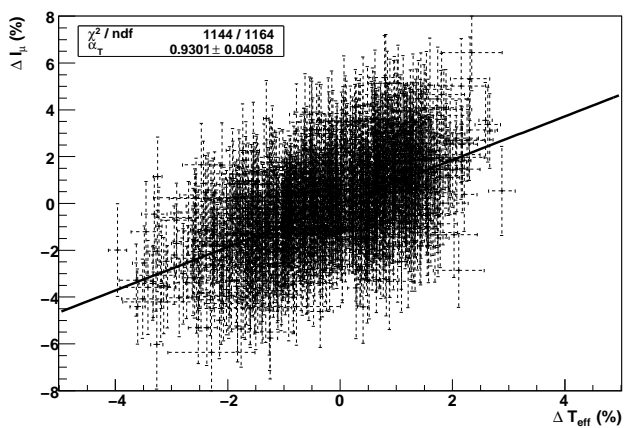}
\caption{$\Delta I_{\mu}/\langle I_{\mu} \rangle$ vs. $\Delta T_{\textrm{eff}}/ \langle T_{\textrm{eff}} \rangle$. Each point is a day.}
\label{fig:correlation}
\end{figure}

\vspace{2mm}

{\bf VII. Correlation.} Observing fig.~\ref{fig:temp_flux} the correlation between fluctuations in atmospheric temperature and cosmic muon flux is evident. 
To quantify such correlation we plotted for every day $\Delta I_{\mu}/\langle I_{\mu} \rangle$ vs $\Delta T_{\textrm{eff}}/\langle T_{\textrm{eff}} \rangle$ in fig. \ref{fig:correlation}. 
Only days with duty cycle $\ge 50\%$ have been included for a total of 1165 days.
The correlation coefficient (R-value) between these two distributions is 0.60 indicating indeed a positive correlation.
To determine $\alpha_T$, a linear regression was performed accounting for error bars on both axes using a numerical minimization method. 
As a result we obtain $\alpha_T = 0.93 \pm 0.04$ with $\chi^2$/NDF = 1144/1164.
We have evaluated the systematic error by varying the assumptions adopted in computing the average flux and temperature with respect to the available data set 
and we have found that it is small compared to the statistical error. 
This result is consistent and features smaller errors when compared  to $\alpha_T = 0.91 \pm 0.07$, the previous measurement by MACRO at Gran Sasso \cite{macro03}. 

\begin{figure}[!thb]
\centering
\includegraphics[width=\columnwidth]{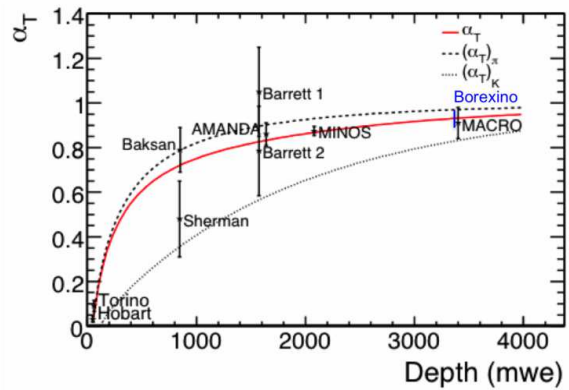}
\caption{Predicted values of $\alpha_T$ as a function of detector depth and the existing measurements at various depths \cite{minos10}.}
\label{fig:minos_alphaT}
\end{figure}
   
The predicted value for $\alpha_{T}$ tends asymptotically to unity with increasing depth of the site, as a deeper rock coverage samples a higher portion of the muon energy spectrum.
Fig. \ref{fig:minos_alphaT} shows this behavior along with existing measurements.
The method to compute predicted values as a function of site depth is detailed in \cite{minos10},  and for LNGS is $\alpha_T=0.92\pm0.02$ considering muon production from both pions and kaons.
The systematic uncertainty was found by modifying the input parameters according to their uncertainties and recalculating. 

With a longer exposure we foresee to measure $\alpha_T$ with better precision and open way to indirect determination of the K/$\pi$ ratio in the interaction of primary cosmic rays in the atmosphere with the method detailed in \cite{minos10, gra10} and probing a complementary energy region compared with existing accelerator experiments.

\vspace{3mm}

{\bf Acknowledgments}

We thank E.W. Grashorn of CCAPP, Ohio State University for insightful discussions and S.M. Osprey of NCAS, University of Oxford (UK) for promptly providing ECMWF air temperature data interpolated on the LNGS coordinates.

\clearpage

\end{document}